\documentclass[fleqn,10pt]{wlscirep}
\usepackage[utf8]{inputenc}
\usepackage[T1]{fontenc}
\usepackage{lineno}
\usepackage{array,multirow,booktabs,tabularx}
\usepackage{amsmath}
\usepackage{algorithm}

\usepackage[noend]{algpseudocode}
\algrenewcommand\algorithmiccomment[1]{\hfill\(\triangleright\) #1}

\title{SpaceTrack-TimeSeries: Time Series Dataset towards Satellite Orbit Analysis}

\author[1,2,3,*,$\dag$]{Zhixin Guo}
\author[1,2,3,*]{Qi Shi}
\author[1,2,3,*,$\dag$]{Xiaofan Xu}
\author[1,2,3]{Sixiang Shan}
\author[1,2,3]{Limin Qin}
\author[1,2,3]{Linqiang Ge}
\author[1,2,3]{Rui Zhang}
\author[1,2,3,4]{Ya Dai}
\author[1,2,3]{Hua Zhu}
\author[1,2,3]{Guowei Jiang}

\affil[1]{Shanghai Satellite Network Research Institute Co., Ltd.,Shanghai, 201210, China}
\affil[2]{State Key Laboratory of Satellite Network, Shanghai, 201210, China}
\affil[3]{Shanghai Key Laboratory of Satellite Network, Shanghai, 201210, China}
\affil[4]{Shanghai Jiao Tong University, Shanghai, 200240, China}
\affil[*]{these authors contributed equally to this work}
\affil[$\dag$]{corresponding author(s): Xiaofan Xu (xiaofanxu@sina.com), Zhixin Guo (stjgzx@alumni.sjtu.edu.cn)}

\begin{abstract}
With the rapid advancement of aerospace technology and the large-scale deployment of low Earth orbit (LEO) satellite constellations, the challenges facing astronomical observations and deep space exploration have become increasingly pronounced. As a result, the demand for high-precision orbital data on space objects—along with comprehensive analyses of satellite positioning, constellation configurations, and deep space satellite dynamics—has grown more urgent. However, there remains a notable lack of publicly accessible, real-world datasets to support research in areas such as space object maneuver behavior prediction and collision risk assessment. This study seeks to address this gap by collecting and curating a representative dataset of maneuvering behavior from Starlink satellites. The dataset integrates Two-Line Element (TLE) catalog data with corresponding high-precision ephemeris data, thereby enabling a more realistic and multidimensional modeling of space object behavior. It provides valuable insights into practical deployment of maneuver detection methods and the evaluation of collision risks in increasingly congested orbital environments.
\end{abstract}

\begin{document}

\flushbottom
\maketitle

\thispagestyle{empty}

\section*{Background \& Summary}

The rapid advancement of aerospace technology, alongside the expansion of the aerospace industry, has considerably accelerated the development of global commercial space initiatives. Prominent companies, including SpaceX, OneWeb, and Amazon, have launched ambitious projects to establish large-scale low Earth orbit (LEO) communication constellations \cite{pachler2021updated}. Since the initial deployment of Starlink satellites, SpaceX has successfully launched over 4,000 satellites \cite{shaengchart2024impact,lagunas2024low}. 

The deployment of mega-constellations entails a significant financial investment, accompanied by extended timelines and inherent uncertainties \cite{hui2025review}. In this case, operators seek to rapidly deploy a partially functional network capable of delivering initial services. For instance, Starlink took over four years to deploy four out of its five planned orbital shells during Phase 1, resulting in the launch of 4,236 satellites, which is short of the original target of 4,408 \cite{ai2025research}. This discrepancy can primarily be attributed to satellite failures and subsequent re-entry, driven by factors such as mechanical malfunctions and geomagnetic storms \cite{grile2025statistical,shirobokov2021survey,caldas2024machine}. Despite the extensive expansion of satellite networks contributing to enhanced global connectivity, it simultaneously exacerbates challenges in astronomical observations and significantly increases the risk of collisions in LEO, thereby posing considerable threats to the long-term sustainability of space operations.

To mitigate the risk of collisions between space objects and enable early maneuver detection, it is essential to obtain precise orbital information. Early research on maneuver detection methods and collision avoidance primarily relied on modeling and simulation based on observational data \cite{vallado2001fundamentals}. Key findings from these studies include the widespread adoption of the Extended Kalman Filter (EKF) \cite{smith1962application,julier1997new,einicke2012robust}, which has become the standard for orbit determination in real-world applications. Furthermore, both analytical solutions \cite{vallado2013improved,miura2009comparison} and numerical approximation methods \cite{urrutxua2016dromo,sharma1988long,aristoff2014orbit,bai2011modified,bradley2012new} have been explored for orbit prediction.

With the development of machine learning and deep learning techniques, data-driven methods have gained increasing attention in this field. To improve the model's representation of reality, researchers have explored approaches that utilize a sufficient number of Two-Line Element (TLE)s as pseudo-observations to fit a high-precision special perturbations numerical propagator \cite{levit2011improved,bennett2012improving,sang2017analytical,san2017hybrid,peng2020machine,muldoon2009improved,peng2018exploring,peng2019comparative,peng2019gaussian,peng2021fusion}. Additionally, some studies focus on mitigating errors introduced by numerical solutions and simplifying linear assumptions \cite{rautalin2017latent,li2021improved,pihlajasalo2018improvement,san2018hybrid,curzi2022two,salleh2021adaptation,li2020machine}.

Both approaches necessitate the acquisition of deep space environmental data to support model adaptation. However, obtaining real-time data on space objects remains a significant challenge. In the case of LEO objects, the sheer number of satellites, coupled with limitations in tracking capabilities—such as equipment availability, observation modes, measurement accuracy, operational capacity, and the spatial distribution of ground stations—renders the collection of high-precision ephemeris for the entire satellite constellation particularly difficult. As a result, real-world datasets in this domain are scarce, leading most experimental studies to rely heavily on simulated or synthetic data. However, such data often fail to capture the complexity and variability of real-world space environments, thereby hindering the direct application of research in maneuver detection methods and collision avoidance, including detector-related studies \cite{tipaldi2022reinforcement}.

In an effort to mitigate this data lacuna, the North American Aerospace Defense Command (NORAD) disseminates cataloged orbital data for space objects via the Space-Track platform, primarily in the form of TLE sets derived from observational measurements \cite{blasch2022space}. However, the inherent low temporal resolution of TLE data renders it suboptimal for accurately characterizing the dynamic trajectories of LEO satellites, particularly those performing frequent orbital maneuvers. Such maneuvers introduce discontinuities into satellite trajectories, thereby substantially degrading the accuracy of TLE-based orbit predictions \cite{liu2024maneuver}. Moreover, TLE data lacks the requisite precision and associated covariance information necessary for rigorously quantifying uncertainties in orbital parameters, thus limiting its utility for applications demanding high precision or operational responsiveness. Addressing the escalating demand for enhanced orbit prediction accuracy, SpaceX has begun disseminating ephemeris data for its Starlink constellation through the same platform. While this ephemeris data offers superior temporal resolution and improved positional fidelity compared to TLE data, it is generated based on pre-defined operational projections rather than real-time telemetry \cite{liu2024maneuver}. Consequently, these predicted trajectories may diverge from the satellites' actual on-orbit behavior and dynamics.

The SpaceTrack-TimeSeries dataset was developed to address the critical gaps identified in existing publicly available orbital data. It offers over 48 weeks of continuous monitoring, combining long-term metrics derived from real-world TLE data with high-resolution Starlink ephemeris. The TLE component is particularly extensive, comprising 329 individual files and totaling 6,989,123 data entries. These records span the period from April 28, 2024, to April 1, 2025, and cover 14,213 unique space objects, including a significant subset of 7,149 Starlink-related satellites. Each TLE record includes essential metadata—such as satellite identifiers, classification levels, timestamps—the six classical Keplerian orbital elements \cite{capderou2005satellite}, and additional parameters critical to orbit propagation and prediction. Complementing this, the Starlink ephemeris component comprises 6,761 files, encompassing a total of 49,853,163 data entries. These records span a shorter but higher-resolution temporal window—from November 25 to December 1, 2024—and exclusively pertain to 6,761 Starlink satellites. Together, the TLE data derived from real-world observations and the ephemeris data based on high-frequency, high-precision predictions provide a complementary and multi-resolution view of orbital dynamics. By integrating both coarse-grained historical records and fine-grained predictive data, the SpaceTrack-TimeSeries dataset offers a realistic and temporally rich representation of the space environment. This enables robust research into spacecraft orbit control, maneuver behavior detection, and collision avoidance strategies in increasingly congested orbital regimes.

\section*{Methods}

The dataset introduced in this study was obtained from Space-Track platform, a research and development initiative under the United States Air Force. Space-Track plays a critical role in promoting global spaceflight safety, preserving the space environment, and supporting the peaceful use of outer space. It achieves this by providing space situational awareness services and disseminating orbital information to a wide array of stakeholders, including satellite operators, academic institutions, and other authorized organizations \cite{blasch2022space}.

\begin{figure*}[!t]
\centering
\includegraphics[scale=0.7]{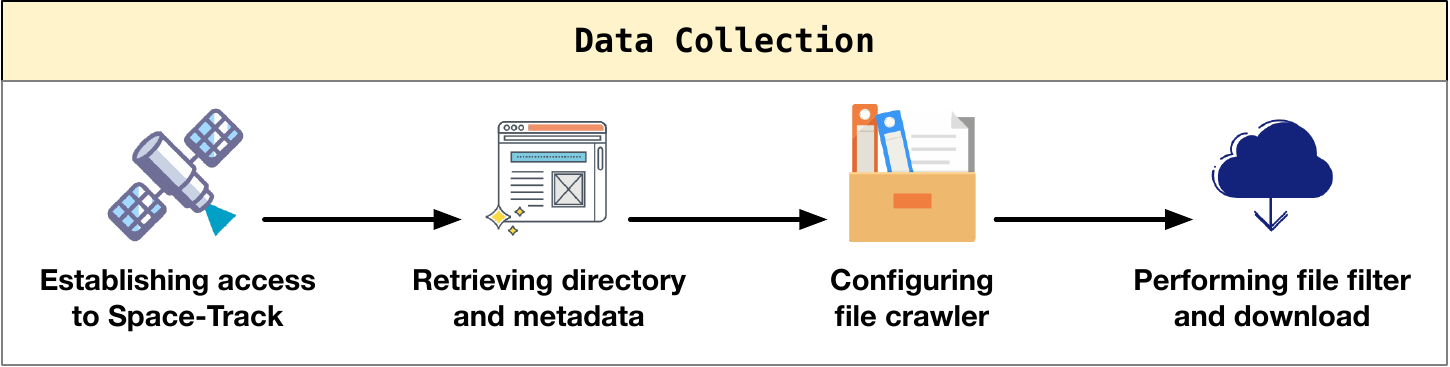}
\caption{An overview of data collection workflow, comprising four main stages: establishing access to Space-Track,  retrieving directory and metadata, configuring file crawling and performing file filtering and download. The data collection process commenced with establishing authenticated access to the Space-Track repository. Subsequently, a systematic extraction of the file directory list and associated metadata was performed. This information informed the configuration of the data crawling protocol, enabling targeted file filtering and subsequent download.}
\label{datacollection}
\end{figure*}

\textbf{Data Collection}. As illustrated in Figure~\ref{datacollection}, the data collection process began with the authentication of programmatic access to the Space-Track data repository. Simultaneously, a POST request was sent to the Space-Track login endpoint to obtain a session cookie, which was subsequently used to enable persistent session access. While data acquisition from Space-Track provides substantial research benefits, it also raises important privacy considerations that necessitate strict compliance with ethical and legal standards. To this end, we first conducted a comprehensive review of the website’s crawler access policies and data usage guidelines. All subsequent data retrieval—restricted to specific datasets such as TLE data and Starlink ephemeris data—was executed in full accordance with these protocols. We explicitly affirm that no data were accessed or processed, whether through automated crawling or manual intervention, in a manner that would violate the site's terms of use.

Following successful authentication, a comprehensive inventory of the directory structure and associated metadata was systematically compiled. Leveraging this information, a customized data crawling protocol was designed and implemented to enable the targeted filtering and retrieval of relevant data files for subsequent analysis. The crawler was configured to process fewer than six files per batch, incorporating a 120-second pause between batches to mitigate the risk of server-side rate limiting. Additionally, a maximum of three retries was permitted for failed requests, with the wait time increasing incrementally by 10 seconds for each subsequent attempt. During the data download phase, stream-based writing was utilized to efficiently manage memory usage and avoid excessive consumption by large files. In instances where an Hypertext Transfer Protocol (HTTP) error response was encountered, a retry mechanism with a dynamically adjusted delay was triggered to address the issue. All other exceptions were systematically logged, and the corresponding files were skipped to maintain continuity and robustness in the data acquisition process.

\begin{figure*}[!t]
\centering
\includegraphics[scale=0.7]{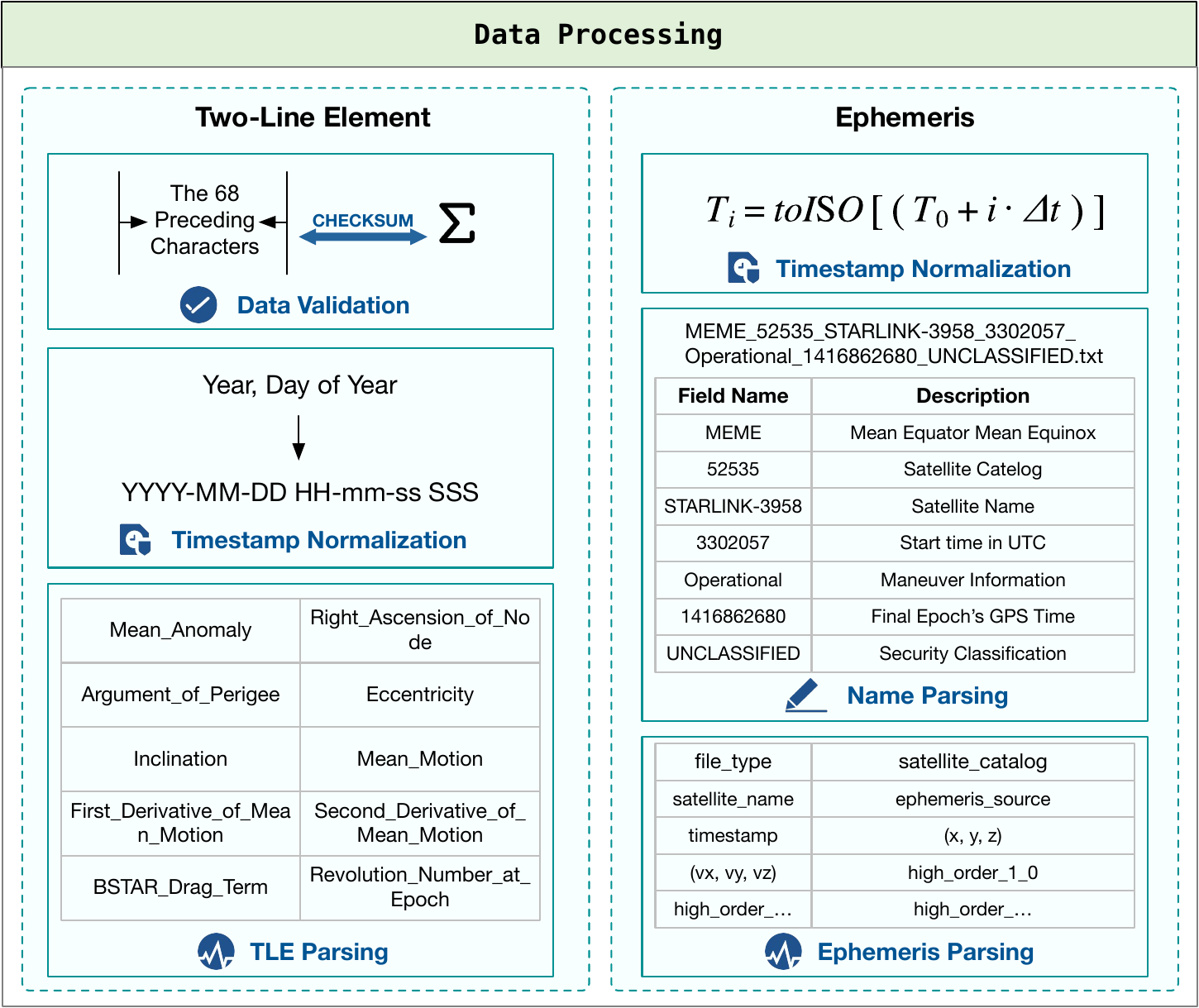}
\caption{An overview of the data processing workflow. The collected TLE and ephemeris datasets are processed using distinct methods tailored to their respective formats. For the TLE data, the primary processing steps include data validation, timestamp normalization, and TLE-specific parsing. In contrast, the processing of ephemeris data involves timestamp normalization, filename-based metadata extraction, and structured ephemeris parsing.}
\label{dataProcessing}
\end{figure*}

\textbf{Data Processing}. To facilitate accurate downstream analysis, the collected orbital datasets must first undergo structured preprocessing. Due to their distinct data structures, the collected TLE and ephemeris datasets require specialized processing methods and workflows, as illustrated in Figure~\ref{dataProcessing}. The processing pipeline for Starlink TLE data comprises three main steps: data validation, timestamp normalization, and parsing adapted to the TLE format. In contrast, the processing of Starlink ephemeris data involves timestamp normalization, metadata extraction from filenames, and parsing procedures tailored to the ephemeris data format.

\textit{TLE data processing.} TLE data, whether transmitted over networks or stored locally, is inherently vulnerable to content corruption arising from signal interference, data degradation, or truncation. To ensure data integrity, we adopt the standard validation procedure recommended by Celestrak \cite{kelso2017challenges}, which requires that the final character of each TLE line serve as a checksum derived from the preceding 68 characters. This approach enables automated verification by a range of software tools and systems. Any mismatch between the computed and recorded checksums indicates data corruption, in which case the affected line is deemed unreliable and is either excluded or flagged. This rigorous validation step is essential for preventing significant inaccuracies in downstream orbital computations.

To facilitate further analysis, we first performed timestamp normalization by converting the original year-and-day-of-year time format into standardized ISO 8601 format \cite{harris2019accurate}. The data was then parsed to extract satellite identifiers, which were subsequently cross-referenced against the comprehensive satellite catalog provided by Space-Track. This filtering process helped eliminate anomalous entries—such as inactive satellites or deep-space debris—thereby reducing interference from irrelevant data sources.

Following the filtering stage, the raw data was converted into a structured format comprising clearly defined orbital parameters, including angular elements, velocity, and eccentricity. To ensure consistency in subsequent scientific computations, all angular measurements were uniformly converted to radians. The original TLE data were retained alongside the structured data to maintain traceability and support debugging when necessary. Each parameter was assigned an unambiguous field name and corresponding physical unit, facilitating seamless integration with data processing tools for batch statistical analysis, modeling, and visualization. In addition, all temporal information was standardized to ISO 8601 format, enabling efficient fusion with multi-source datasets and streamlining time-series analysis. The resulting structured dataset is well-suited not only for research-oriented orbital analysis but also for practical engineering applications, such as orbital dynamics simulations and satellite collision avoidance systems.

\textit{Ephemeris data processing.} Timestamp normalization is likewise crucial for Starlink ephemeris data to ensure consistency in downstream analyses. In the source files, temporal information is implicitly encoded via a defined ephemeris start time and a constant temporal step between consecutive data points. Explicit timestamps were therefore reconstructed by systematically iterating through each dataset, assigning a distinct time value to every data point. These timestamps were subsequently formatted and stored in compliance with the ISO 8601 date-time standard.

In contrast to TLE data, Starlink ephemeris filenames directly incorporate a significant volume of metadata. Adopting the approach proposed in previous studies, this embedded metadata was systematically extracted via filename parsing, a procedure illustrated in Figure~\ref{dataProcessing} under “Name Parsing”.

For robust preprocessing and further analytical applications, each file was converted into a structured representation. This representation ensures each record contains essential orbital analysis parameters, including: satellite catalog number, satellite name, data provenance, the ISO-formatted timestamp, three-dimensional position and velocity vectors, and higher-order perturbation terms.

\begin{figure*}[!t]
\centering
\includegraphics[scale=0.7]{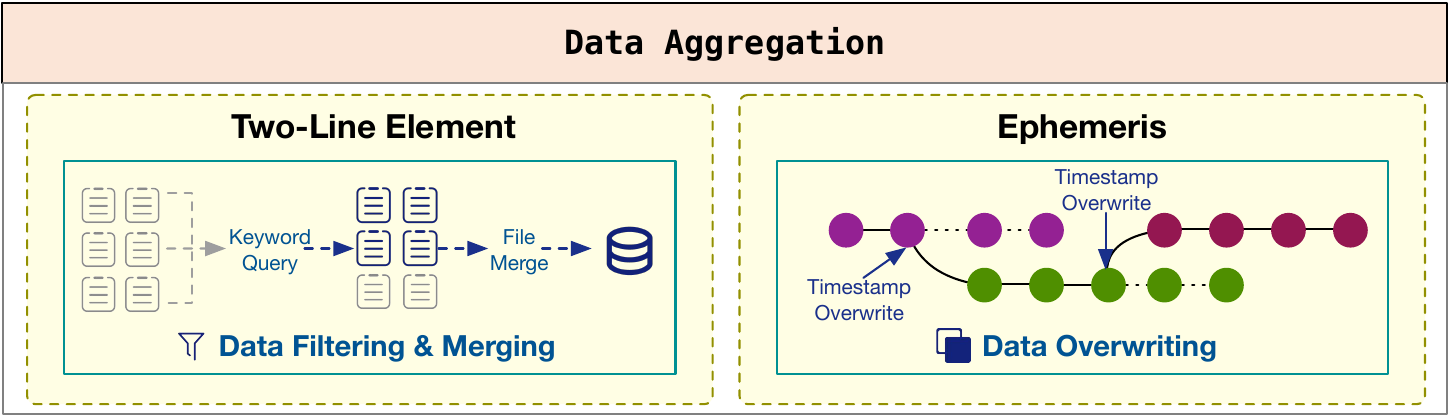}
\caption{Overview of the data aggregation workflow. Distinct aggregation workflows are applied to processed TLE and ephemeris datasets, reflecting their inherent protocol differences. TLE data aggregation involves systematic filtering and merging of entries to construct a unified and continuous dataset. In contrast, ephemeris data aggregation employs a data overwriting strategy, wherein overlapping entries are replaced based on temporal resolution and data source priority to ensure consistency and accuracy.}
\label{dataaggregation}
\end{figure*}

\textbf{Data Aggregation}. The data aggregation workflow is illustrated in Figure~\ref{dataaggregation}. The processed TLE and ephemeris datasets are aggregated using distinct methodologies, each specifically adapted to the structural characteristics and storage formats of the respective datasets. The aggregation of TLE data involves a two-step process: first, a query-based filtering procedure is applied to identify and select candidate satellites belonging to the Starlink constellation; this is followed by the merging of all relevant data files associated with each unique satellite.

In contrast, ephemeris data aggregation primarily utilizes a data overwriting strategy. This approach is informed by the Starlink ephemeris data publishing protocol, which specifies that these files contain orbital predictions derived from satellite trajectory planning models. Because the predictive accuracy of such files degrades over time, and to ensure both real-time precision and accommodation for adjustments in orbital strategies, Starlink issues frequent data updates. These updates often result in temporal overlaps between newly released and previously published ephemeris files. Consequently, during the aggregation process, these temporal overlaps are resolved by systematically replacing earlier data points with the most recently published values, thereby ensuring the aggregated dataset reflects the most current and accurate orbital information available.

\section*{Data Records}


All data curated and presented in this study are publicly accessible via Figshare repository \cite{guo2025spacetrack}. The overall structure of the SpaceTrack-TimeSeries dataset primarily comprises two key components: ephemeris data and TLE data. The ephemeris data component, encompassing essential published metrics, is provided in Comma-Separated Values (CSV) format, wherein each file is systematically named according to its associated metadata. Specifically, this naming convention structures each filename from distinct components—identified as three primary metadata fields and one data processing identifier—which are concatenated using underscores (\_). These metadata fields are the Orbital Reference System Identifier (ORSI), the satellite catalog number, and the satellite name.

The internal data fields within these ephemeris CSV files are comprehensively described in Table~\ref{ephemeris_table}. Each data record incorporates a set of key identifiers: the Coordinate Reference System; the Satellite\_Number, representing the unique sequential identifier from the United Space Command (USSPACECOM); the human-readable Satellite\_Name; and the Ephemeris\_Source, which is uniformly set to \texttt{'blend'} in these files to signify a composite product. Additionally, every entry includes a precise Timestamp, the satellite’s three-dimensional position (X, Y, Z) and velocity (Vx, Vy, Vz) vectors. Crucially, the record also contains an extensive set of High\_Order\_X\_Y parameters. These parameters represent the covariance matrix elements of the satellite’s state vector, quantifying the uncertainties associated with orbit determination results—information essential for various advanced applications.

\begin{table}[htbp]
\renewcommand{\arraystretch}{1.3}
\begin{center}
\begin{tabularx}{\textwidth}{>{\raggedright\arraybackslash}p{3.5cm}|>{\raggedright\arraybackslash}X}
\hline
\textbf{Column Name} & \textbf{Description} \\
\hline
Coordinate & The specific coordinate reference system employed for the ephemeris data. \\
\hline
Satellite\_Number & The unique sequential number allocated by USSPACECOM. \\
\hline
Satellite\_Name & A human-readable designation for on-orbit artificial objects, including operational satellites, defunct satellites, rocket bodies, and larger debris fragments. \\
\hline
Ephemeris\_Source & The ephemeris data source—as indicated by the uniform designation \texttt{'blend'} in the released files—denotes that the data is a composite product. \\
\hline
Timestamp & The precise temporal reference for the data contained in each row, with particular relevance to its state vector elements such as position and velocity. \\
\hline
X, Y, Z & The X/Y/Z components of the satellite’s position vector in the designated coordinate system. \\
\hline
Vx, Vy, Vz & The X/Y/Z components of the satellite’s velocity vector in the coordinate system. \\
\hline
High\_Order\_1\_0 & \multirow{5}{=}{Covariance matrix elements of the satellite’s state vector quantify the uncertainties associated with orbit determination results. This information is essential for applications such as evaluating orbit accuracy, computing collision probabilities, planning maneuvers, optimizing sensor tasking, and enabling data fusion.} \\
\cline{1-1}
High\_Order\_1\_1 & \\
\cline{1-1}
$\dots$ & \\
\cline{1-1}
High\_Order\_3\_5 & \\
\cline{1-1}
High\_Order\_3\_6 & \\
\hline
\end{tabularx}
\end{center}
\caption{Descriptions of columns in the satellite ephemeris dataset.}
\label{ephemeris_table}
\end{table}

The TLE data component is structured into two distinct subsets: full\_data and overlapped\_data. The full\_data subset provides comprehensive TLE records for all collected satellites. The overlapped\_data subset, designed for compatibility with the ephemeris dataset, contains only TLE records for Starlink satellites spanning the period from November 23 to November 26, 2024—aligned with the temporal coverage of the corresponding ephemeris data. Both the full\_data and overlapped\_data subsets are delivered in CSV, with filenames assigned according to a time series identifier.


All TLE data files across the aforementioned subsets adhere to a standardized columnar schema, with detailed field descriptions provided in Table~\ref{tle_table}. Collectively, these fields furnish a comprehensive characterization of a satellite’s orbital state at a specified epoch. Key identification parameters include the Satellite\_Number (a unique sequential identifier assigned by USSPACECOM), the human-readable Satellite\_Name, and the International\_Designator (COSPAR ID), which together ensure precise and standardized referencing of artificial space objects. The Class field denotes the security classification of the TLE data—typically \texttt{'U'} (Unclassified) for publicly released records—while the Timestamp field specifies the exact epoch for which the associated orbital elements are valid.

At the core of each TLE record are the six classical Keplerian orbital elements, which collectively define the satellite’s orbital geometry and its position within that framework. These parameters—including Inclination (defining orbital tilt), Right\_Ascension\_of\_Node (RAAN, orienting the orbit in inertial space), Eccentricity (quantifying deviation from a circular orbit), Argument\_of\_Perigee (orienting the ellipse within its orbital plane), Mean\_Anomaly (parameterizing position relative to perigee), and Mean\_Motion (average daily orbital revolutions)—are fundamental for orbit determination and trajectory propagation. To support the modeling of orbital evolution, particularly for use with SGP4/SDP4 propagation algorithms \cite{dong2010accuracy}, dedicated parameters addressing perturbations are also included. The First\_Derivative\_of\_Mean\_Motion captures the temporal rate of change of the Mean Motion, while the Second\_Derivative\_of\_Mean\_Motion represents the acceleration or deceleration of this rate. The BSTAR\_Drag\_Term quantifies atmospheric drag effects and is integral to accurate orbital propagation using these models.

To facilitate operational tracking and data versioning, the dataset includes the Revolution\_Number\_at\_Epoch—denoting the cumulative number of complete orbits up to the specified epoch—and the Element\_Number. The Element\_Number acts as a sequential identifier, enabling differentiation between multiple TLE sets for a single satellite that correspond to different epochs, a feature especially valuable for determining data recency.

\begin{table}[htbp]
\renewcommand{\arraystretch}{1.3}  
\begin{center}
\begin{tabularx}{\textwidth}{>{\raggedright\arraybackslash}p{4.5cm}|>{\raggedright\arraybackslash}X}
\hline
\textbf{Column Name} & \textbf{Description} \\
\hline
Satellite\_Number & The unique sequential number allocated by United States Space Command (USSPACECOM). \\
\hline
Satellite\_Name & A human-readable designation for on-orbit artificial objects, including operational satellites, defunct satellites, rocket bodies, and larger debris fragments. \\
\hline
International\_Designator & A standardized system for uniquely identifying artificial space objects, including satellites, rocket bodies, and debris, maintained by the Committee on Space Research (COSPAR). \\
\hline
Class & The security classification level of an orbital element set, with common values being \texttt{'U'} (Unclassified), \texttt{'C'} (Classified), and \texttt{'S'} (Secret). \\
\hline
Timestamp & The specific moment in time for which a given set of orbital elements is precisely defined. \\
\hline
Mean\_Anomaly & One of the six Keplerian orbital elements, representing an angle that defines a satellite's position in its elliptical orbit relative to perigee through a conceptual simplification. \\
\hline
Right\_Ascension\_of\_Node & One of the six classical Keplerian orbital elements, defines the orientation of a satellite's orbital plane in inertial space. \\
\hline
Argument\_of\_Perigee & One of the six classical Keplerian orbital elements, defines the orientation of an orbiting body's major axis (specifically, the direction of perigee) within its orbital plane. \\
\hline
Eccentricity & One of the six classical Keplerian orbital elements, is a dimensionless parameter that quantifies the degree to which an orbit's shape deviates from that of a perfect circle. \\
\hline
Inclination & One of the six classical Keplerian orbital elements. It defines the angular tilt of an orbiting body's orbital plane relative to a fundamental reference plane, which for Earth-orbiting satellites is the Earth's equatorial plane. \\
\hline
Mean\_Motion & Denotes the average number of complete revolutions (or orbits) an Earth-orbiting satellite executes per day. \\
\hline
First\_Derivative\_of \_Mean\_Motion & The first derivative of Mean Motion specifies its temporal rate of change, consistent with the fundamental mathematical concept that a first derivative indicates the rate at which one quantity varies relative to another (usually time). \\
\hline
Second\_Derivative\_of \_Mean\_Motion & The second derivative of Mean Motion defines the rate of change of the first derivative of Mean Motion, effectively signifying the \texttt{'acceleration'} or \texttt{'deceleration'} of the Mean Motion's rate of change. \\
\hline
BSTAR\_Drag\_Term & A parameter quantifying atmospheric drag, which is a fundamental component of the SGP4/SDP4 orbit propagation models. These are standard tools for predicting artificial Earth satellite orbits and are expressly designed for use with TLE. \\
\hline
Revolution\_Number\_at\_Epoch & The cumulative number of complete revolutions (or orbits) performed by a satellite from its launch (or an alternative designated start point) up to the TLE epoch time. \\
\hline
Element\_Number & A sequential identifier assigned to each TLE set of a particular satellite, primarily to distinguish between multiple TLEs for that satellite which correspond to different epochs. \\
\hline
\end{tabularx}
\end{center}
\caption{Descriptions of columns in the TLE dataset.}
\label{tle_table}
\end{table}

\section*{Technical Validation}

\textbf{TLE Distributional Integrity Validation}.
\begin{figure*}[!t]
\centering
\includegraphics[scale=0.51]{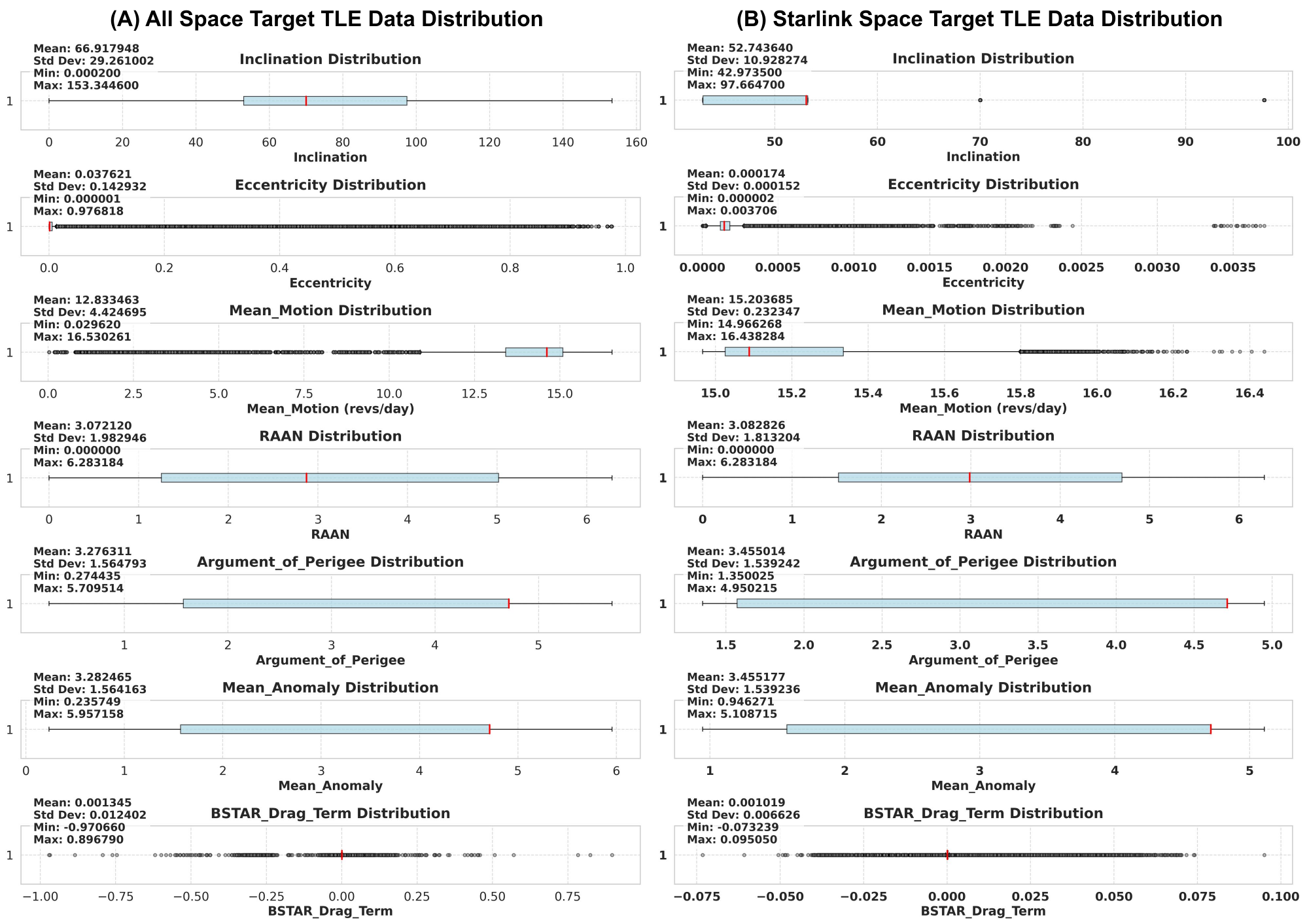}
\caption{Distribution integrity validation of orbital parameters within TLE dataset. Panel (A) illustrates the distribution of all space object TLE data, while panel (B) focuses specifically on Starlink satellites. Both panels depict the frequency distributions of key orbital parameters, including inclination, eccentricity, mean motion, right ascension of ascending node, argument of perigee, mean anomaly, and the BSTAR drag term.}
\label{tle-orbital-parameter}
\end{figure*}

To evaluate the distributional integrity of TLE data, a comprehensive statistical analysis was conducted on seven key orbital parameters: inclination, eccentricity, mean motion, right ascension of the ascending node (RAAN), argument of perigee, mean anomaly, and the BSTAR drag term. This analysis encompassed two distinct subsets: the complete catalog of tracked space objects and a focused subset comprising only Starlink satellites.

As depicted in Figure~\ref{tle-orbital-parameter} (A), the box plots for the general TLE dataset reveal substantial heterogeneity in orbital characteristics. This is consistent with the diverse nature of the tracked objects, which include operational satellites, decommissioned spacecraft, rocket bodies, and orbital debris. Inclination values span a broad range (mean=$66.917^\circ$, standard deviation=$29.261^\circ$, min=$0.000^\circ$, max=$153.345^\circ$), covering near-equatorial to polar and even retrograde orbits. Eccentricities vary from nearly circular to highly elliptical (mean=$0.038^\circ$, standard deviation=$0.143^\circ$, min=$0.000^\circ$, max=$0.977^\circ$), reflecting the wide variety of mission types and orbital regimes represented in the global TLE catalog, such as LEO, Medium Earth Orbit (MEO), Geostationary Earth Orbit (GEO), and Highly Elliptical Orbit (HEO). Figure~\ref{tle-orbital-parameter} (A) also shows substantial dispersion across all other parameters. For instance, RAAN values span nearly the full [0,$2\pi$] range (mean=3.072 radians, standard deviation=1.983 radians, min=0.000 radians, max=6.283 radians), indicating a lack of structured orbital plane alignment. Similarly, both the argument of perigee (mean=3.276 radians, standard deviation=1.565 radians) and the mean anomaly (mean=3.282 radians, standard deviation=1.564 radians) display broad distributions, characteristic of the random orbital phasing often associated with derelict objects. The BSTAR drag term, which encapsulates the combined effects of atmospheric density, satellite cross-sectional area, and mass, also shows a wide spread (mean=0.001345 $R_E^{-1}$, standard deviation=0.012402 $R_E^{-1}$, min=-0.970660 $R_E^{-1}$, max=0.896790 $R_E^{-1}$), including negative values—potentially indicating data anomalies or legacy entries. (Note: Units for BSTAR are typically Earth radii -1, added for context, assuming standard TLE definition \cite{lanz2007grid}). Mean motion for the general TLE dataset also shows considerable spread (mean=12.833 \text{revs/day}, standard deviation=4.425 \text{revs/day}, min=0.030 \text{revs/day}, max=16.530 \text{revs/day}).

In contrast, Figure~\ref{tle-orbital-parameter} (B) illustrates the markedly uniform distributions observed in the Starlink subset, highlighting the constellation’s systematic and coordinated deployment strategy. Inclinations in the Starlink subset have a mean of $52.744^\circ$ and a standard deviation of $10.928^\circ$, with values ranging from $42.974^\circ$ to $97.665^\circ$ . This distribution reflects deployment into multiple specific orbital shells (including major clusters around $53^\circ$, with the range extending up to $97.665^\circ$ for other shells) designed by SpaceX to achieve near-global coverage. Eccentricities are consistently close to zero (mean=0.000174, std=0.000152, min=0.000002, max=0.003706), consistent with nearly circular orbits optimized for maintaining stable ground tracks. Mean motion values exhibit minimal variation (mean=15.203 \text{revs/day}, standard deviation=0.232 \text{revs/day}, min=14.966 \text{revs/day}, max=16.438 \text{revs/day}), further reflecting uniform orbital altitudes. The RAAN distribution within the Starlink subset (mean=3.083 radians, standard deviation=1.813 radians, min=0.000 radians, max=6.283 radians) indicates intentional plane spacing across the full range of nodal longitudes. The argument of perigee (mean=3.455 radians, standard deviation=1.539 radians) and mean anomaly (mean=3.455 radians, standard deviation=1.539 radians) show less variability than in the general TLE dataset, although they still span a moderate range. Notably, BSTAR values are tightly bounded (mean=0.001019 $R_E^{-1}$, standard deviation=0.006626 $R_E^{-1}$, min=-0.073239 $R_E^{-1}$, max=0.095050 $R_E^{-1}$); this narrow range reflects consistent atmospheric drag characteristics attributable to standardized satellite designs and tightly controlled orbital altitudes. Collectively, these results underscore a sharp contrast between the structured, homogeneous nature of the Starlink constellation and the heterogeneous, uncoordinated composition of the broader cataloged space object population.

\begin{figure*}[!t]
\centering
\includegraphics[scale=0.34]{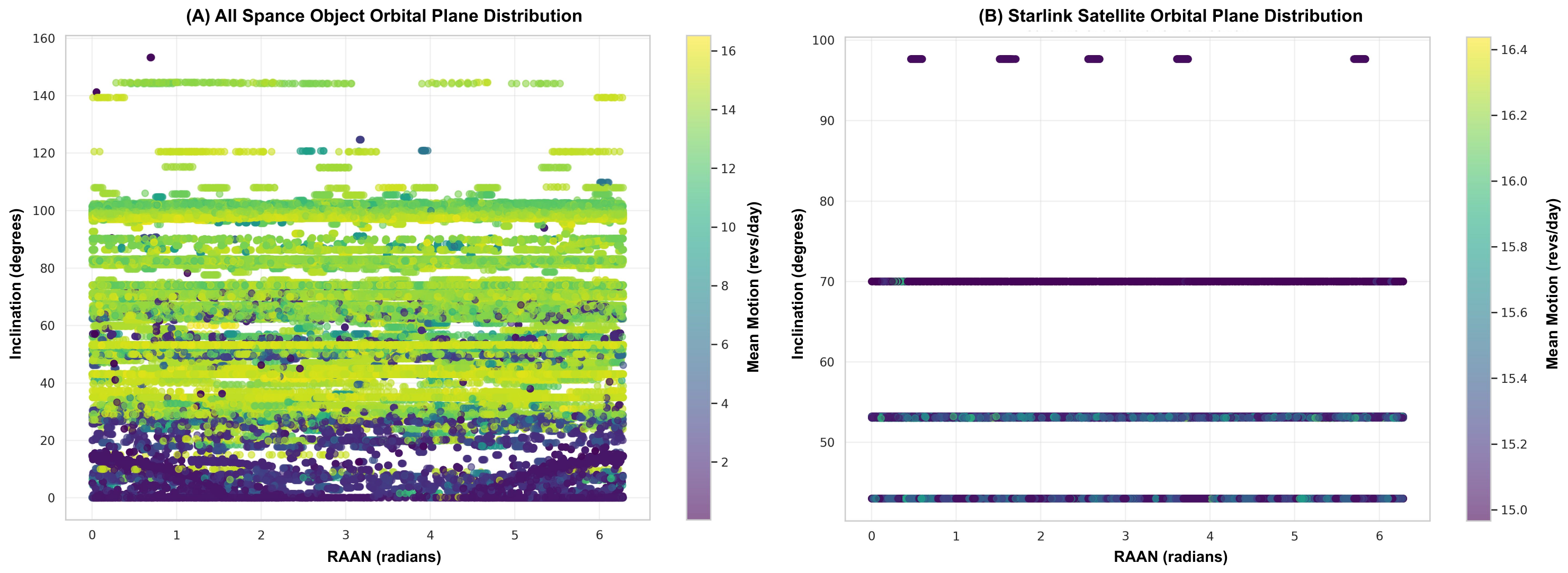}
\caption{Distribution integrity validation of orbital plane within TLE dataset. Panel (A) presents the spatial distribution of orbital planes for all cataloged space objects based on TLE data. Panel (B) shows the orbital plane layout for Starlink satellites, highlighting their structured and dense deployment within specific inclination shells and right ascension of ascending node intervals.}
\label{tle-orbit-plane}
\end{figure*}

Figure~\ref{tle-orbit-plane} further corroborates the distributional patterns observed in the box plots through a three-dimensional visualization of orbital plane parameters. Figure~\ref{tle-orbit-plane} (A) depicts the spatial configuration of all cataloged space objects in a parameter space defined by inclination, right ascension of ascending node, and mean motion. Consistent with the results in Figure~\ref{tle-orbital-parameter} (A), the distribution reveals pronounced heterogeneity: inclination values span from near-equatorial to polar and retrograde orbits, right ascension of ascending node values are scattered broadly without discernible structure, and mean motion varies significantly, reflecting the presence of objects across multiple orbital regimes (LEO, MEO, GEO, HEO). This disordered spatial arrangement reinforces the interpretation of a diverse and uncoordinated population composed of satellites, rocket bodies, and debris.

In contrast, Figure~\ref{tle-orbit-plane} (B) presents the orbital plane distribution of Starlink satellites, exhibiting a highly regular and symmetric structure that echoes the uniformity. Inclinations form a tight vertical band, confirming the use of a shell-based deployment strategy. right ascension of ascending node values demonstrate a quasi-periodic spacing pattern, indicative of intentional plane distribution for coverage optimization and conjunction risk mitigation. Mean motion is confined to a narrow range, reaffirming the constellation's uniform altitudinal configuration. Together, these visualizations illustrate the stark contrast between uncoordinated legacy objects and the geometrically disciplined layout of modern satellite constellations.

\textbf{Ephemeris Distributional Integrity Validation}.

To facilitate an unbiased, one-to-one comparison between the high-cadence ephemeris data and the low-cadence TLE sets, the uniformly sampled instantaneous osculating Keplerian elements \cite{peloni2018osculating} are transformed into 24-hour Brouwer-mean elements \cite{marsouin1991navigation} using Algorithm~\ref{alg:osc2Bmean}.  
Given a sampling step of $\Delta t_{\text{step}}$ and a fixed window width of $\Delta t_{\text{win}} = 24$ H, the algorithm first defines the half-window length
\begin{equation}
m = \left\lfloor \frac{\Delta t_{\text{win}}}{\Delta t_{\text{step}}} \right\rfloor ,
\label{eq:half-window}
\end{equation}
which produces a rectangular moving window comprising $(2m+1)$ samples.  
For each epoch~$k$, the osculating six-tuple $\{a_k,e_k,i_k,\Omega_k,\omega_k,M_k\}$ is processed in two sequential steps:

\begin{enumerate}
  \item \textbf{Phase-continuity correction.}  
        The angular variables $\Omega_k$, $\omega_k$, and $M_k$ are unwrapped,  
        $\Omega_k,\omega_k,M_k \leftarrow \operatorname{unwrap}(\cdot)$,  
        to eliminate spurious $2\pi$ discontinuities.
  \item \textbf{Low-pass filtering.}  
        Each element $p \in \{a,e,i,\Omega,\omega,M\}$ is subjected to a box-car average,
        \begin{equation}
            \bar{p}_k \;=\; \frac{1}{2m+1}\sum_{j=-m}^{m} p_{k+j},
        \end{equation}
       thereby attenuating all periodic signals (or variations/components/cycles) with periods shorter than 24 H..
\end{enumerate}

The filtered angular components are subsequently wrapped back to the principal interval,  
$(\bar{\Omega}_k,\bar{\omega}_k,\bar{M}_k) \bmod 2\pi \in [0,2\pi)$.  
The resulting sequence
$\{\bar a_k,\bar e_k,\bar i_k,\bar\Omega_k,\bar\omega_k,\bar M_k\}$  
is devoid of short-period perturbations and fully consistent with the Brouwer analytic mean model underlying TLEs.  
Accordingly, the ephemeris-derived and TLE-derived element sets share an identical dynamical reference, which enables rigorous cross-validation, systematic discrepancy analysis, and subsequent orbit-prediction error studies within a unified metric framework.

\begin{algorithm}[H]
\caption{Osculating Series $\;\Rightarrow\;$ 24 H Brouwer-Mean Series}
\label{alg:osc2Bmean}
\begin{algorithmic}[1]
  \Require $\bigl\{t_k,a_k,e_k,i_k,\Omega_k,\omega_k,M_k\bigr\}_{k=1}^{N}$ 
           \hfill$\triangleright$ uniformly–sampled \emph{osculating} elements  
  \Require Sampling step $\Delta t_{\text{step}}$ (e.g.\ 10 min); window width $\Delta t_{\text{win}}=24$ h  
  \Ensure  $\bigl\{\bar a_k,\bar e_k,\bar i_k,\bar\Omega_k,\bar\omega_k,\bar M_k\bigr\}$  
           \hfill$\triangleright$ 24 h Brouwer-mean elements  
  \Statex \textbf{Per-epoch osculating six-tuple}  
  \Statex \quad $a_k$ — semi-major axis  
  \Statex \quad $e_k$ — eccentricity  
  \Statex \quad $i_k$ — inclination (rad)  
  \Statex \quad $\Omega_k$ — right ascension of ascending node (RAAN, rad)  
  \Statex \quad $\omega_k$ — argument of perigee (rad)  
  \Statex \quad $M_k$ — mean anomaly (rad)  
  \Statex \textbf{Corresponding 24 h means}  
  \Statex \quad $\bar a_k,\bar e_k,\bar i_k,\bar\Omega_k,\bar\omega_k,\bar M_k$ — same quantities after low-pass filtering  
  \Statex \textbf{Window parameter} $m=\lfloor\Delta t_{\text{win}}/\Delta t_{\text{step}}\rfloor$ — half-window length (samples)  
  \State $m\gets\lfloor\Delta t_{\text{win}}/\Delta t_{\text{step}}\rfloor$
  \For{\textbf{each} element $p\in\{a,e,i,\Omega,\omega,M\}$}
     \If{$p$ is angular ($\Omega,\omega,M$)}
        \State $p_k\gets\operatorname{unwrap}(p_k)$   \hfill$\triangleright$ remove $2\pi$ jumps
     \EndIf
     \State $\displaystyle\bar p_k\gets\frac{1}{2m+1}\sum_{j=-m}^{m}p_{k+j}$ \hfill$\triangleright$ box-car average
  \EndFor
  \State $(\bar\Omega_k,\bar\omega_k,\bar M_k)\gets\bmod 2\pi$ into $[0,2\pi)$
  \State \Return $\bigl(\bar a_k,\bar e_k,\bar i_k,\bar\Omega_k,\bar\omega_k,\bar M_k\bigr)$
\end{algorithmic}
\end{algorithm}

\begin{figure*}[!t]
\centering
\includegraphics[scale=0.45]{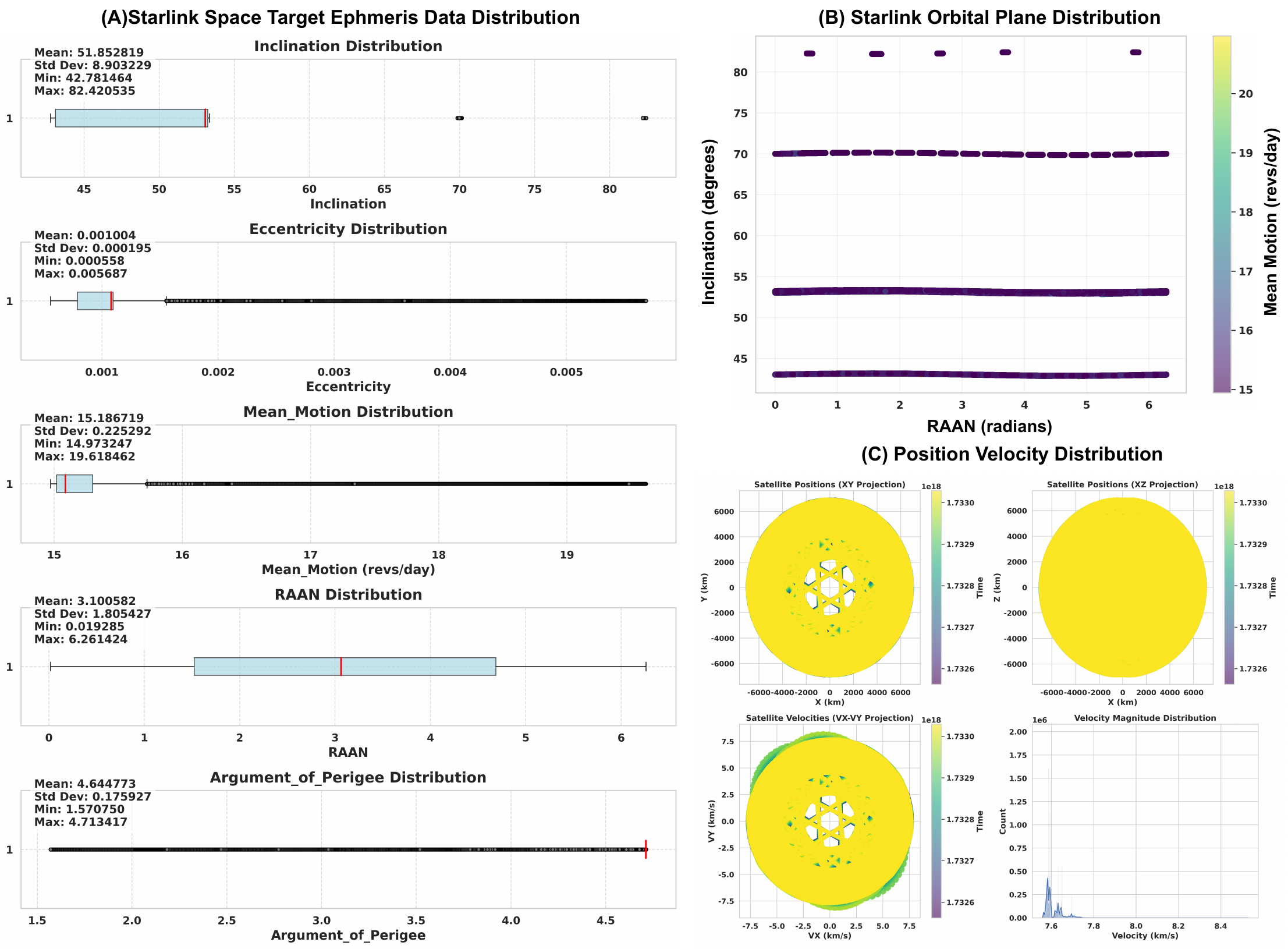}
\caption{Distribution integrity validation of the Starlink ephemeris dataset.  Panel~(A) presents box-plots of five key orbital elements—inclination, eccentricity, mean motion, RAAN, and argument of perigee—demonstrating their statistical consistency with the constellation’s shell-based design. Panel (B) shows the distribution of orbital planes, while Panel (C) illustrates the corresponding position–velocity distribution.}
\label{ephemeris-boxplots}
\end{figure*}

Figure~\ref{ephemeris-boxplots} (A) illustrates the distribution of five key orbital parameters—inclination, eccentricity, mean motion, right ascension of ascending node, and argument of perigee—extracted from the Starlink ephemeris dataset. As visualized through box plots, these parameters exhibit varying degrees of concentration and spread, reflecting both the systematic deployment strategy and minor natural variations within the constellation. The inclination values are centered around a mean of 51.85 degree with a standard deviation of 8.90 degree, ranging from 42.78 degree to 82.42 degree. While most satellites cluster near the nominal orbital inclination of 53 degree, the presence of higher inclination outliers indicates the existence of additional polar or near-polar deployment shells. The eccentricity values are tightly distributed, with a mean of 0.001004 and a small standard deviation of 0.000195, suggesting that Starlink satellites generally operate in near-circular orbits, as expected for constellations requiring stable ground track coverage. The mean motion, expressed in revolutions per second, is highly consistent across the constellation, with a mean of 15.187 \text{revs/day} and minimal variability (standard deviation = 0.225292), corresponding to uniform orbital altitudes. The RAAN values show a broad spread (standard deviation = 1.805), despite a near-zero mean, indicating a wide distribution of orbital planes. This distribution aligns with the constellation’s design goal of achieving global coverage through deliberate spacing of orbital planes around the Earth. Lastly, the argument of perigee displays moderate dispersion (mean = 4.645, standard deviation = 0.176), with values ranging from 1.571 to 4.713. While this parameter has limited operational impact for near-circular orbits, its observed variability reflects the dynamic evolution of orbital geometry due to natural perturbations. Overall, the statistical distributions validate the internal consistency of the Starlink ephemeris data and provide empirical evidence of its carefully orchestrated yet physically realistic deployment pattern.

Figure~\ref{ephemeris-boxplots} (B) delineates the distribution of orbital planes in a three-dimensional parameter space spanned by inclination, right ascension of the ascending node (RAAN), and mean motion.  Mirroring the pattern observed in Figure~\ref{tle-orbit-plane} (B), the satellites cluster into distinct, vertically aligned bands at common inclinations, thereby reaffirming the shell-based deployment strategy adopted for the constellation.  The RAAN values follow a structured, quasi-periodic sequence, reflecting the deliberate spacing of orbital planes that secures near-global coverage while mitigating in-plane conjunction risk.  The extremely narrow spread in mean motions further indicates operation at essentially uniform orbital altitudes, underscoring the constellation’s highly coordinated architecture. Figure~\ref{ephemeris-boxplots} (C) portrays the instantaneous position and velocity vectors projected into Cartesian space $(X, Y, Z)$~\cite{ude2014orientation}.  The resulting structure forms a smooth, toroidal envelope, confirming that the satellites are homogeneously distributed both spatially and kinematically.  The absence of significant clustering or outliers implies effective orbit maintenance and precise station-keeping. Taken together, Figure~\ref{ephemeris-boxplots} (B) and (C) underscore the geometric and dynamical regularity that characterizes the Starlink constellation.  The coherent structure evident in both orbital-element space and state-vector space attests to the internal consistency of the ephemeris dataset and validates its suitability for demanding applications such as high-fidelity orbit propagation, target tracking, and large-scale constellation modeling.

\textbf{Overall Consistency Validation}.

\begin{figure*}[!t]
\centering
\includegraphics[scale=0.47]{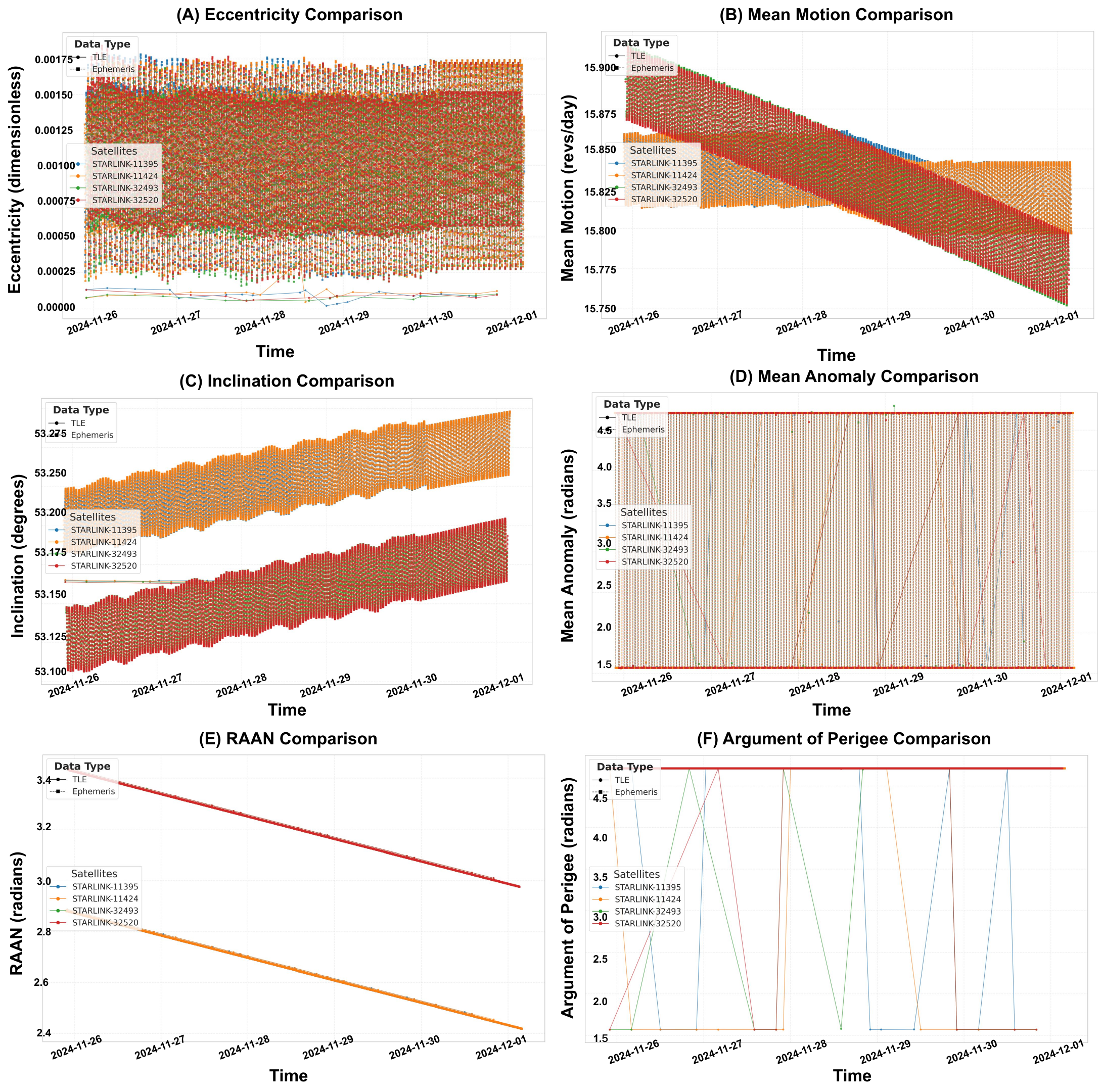}
\caption{Consistency validation between TLE and ephemeris data for four randomly selected Starlink satellites.  Panels: (A) eccentricity, (B) mean motion, (C) inclination, (D) mean anomaly, (E) right ascension of the ascending node (RAAN), and (F) argument of perigee.  
Circular solid markers represent TLE values, whereas rectangular dashed markers denote ephemeris values.  Color coding of satellites: blue demonstrates \textsc{STARLINK-11395}; orange demonstrates \textsc{STARLINK-11424}; green demonstrates \textsc{STARLINK-32493}; red demonstrates \textsc{STARLINK-32520}.
}
\label{comparison}
\end{figure*}

To quantify the mutual fidelity of TLE and ephemeris representations, this study tracked six fundamental orbital elements for a randomly selected quartet of Starlink spacecraft, with their temporal evolutions plotted in Figure~\ref{comparison}. Panels (A)–(F) therein provide a comparative analysis of eccentricity, mean motion, inclination, mean anomaly, RAAN, and argument of perigee, respectively. Within these plots, TLE-derived estimates are indicated by circular solid markers, while ephemeris-derived estimates are denoted by rectangular dashed markers; the four distinct spacecraft are colour-coded for enhanced clarity.

Figure~\ref{comparison} reveals that TLE sets and ephemerids exhibit close agreement at the macroscopic level, jointly capturing the essential characteristics of the satellite orbits. However, pronounced discrepancies emerge upon consideration of fine-scale details, with these differences fundamentally rooted in the distinct design philosophies underpinning the two datasets. TLEs encode Brouwer-mean orbital elements, specifically tuned for analytical propagators like SGP4/SDP4, and are designed to intentionally filter out short-period perturbations to facilitate robust long-term forecasting. Ephemerids, in stark contrast, furnish instantaneous, high-resolution state vectors that preserve high-frequency oscillations and other physical disturbances, thereby enabling centimeter-level precision for tasks such as precise orbit determination, conjunction analysis, and detailed scientific investigation. Moreover, the temporal cadence of ephemerids is orders of magnitude denser than that of TLEs, allowing for the resolution of subtle dynamical features that the sparsely sampled TLE catalog cannot reveal. Consequently, the observed fine-scale mismatches are not indicative of errors but are rather inherent reflections of each dataset's distinct modeling assumptions and operational objectives. Acknowledging these provenance-dependent differences is therefore crucial not only for accurately interpreting numerical divergences but also for informing the application of suitable methodologies to normalize or integrate these distinct data types within an appropriate analytical framework tailored to research objectives. Such an approach enhances research efficacy by leveraging the complementary insights from these multi-dimensional data sources, while fostering a deeper appreciation for the underlying consistency both formats maintain in depicting broad orbital behavior.

\section*{Usage Notes}

The SpaceTrack-TimeSeries dataset, encompassing both TLE and ephemeris data types, offers distinct components suitable for independent application in satellite orbit research. For methodological clarity and rigor, users are required to precisely specify the dataset component (TLE or ephemeris, including any specific aggregation levels) employed in their analysis, ensuring this choice is congruent with their defined research objectives. Furthermore, in instances where multiple dataset types or aggregation levels are utilized, findings must be reported discretely for each configuration—avoiding conflation—to uphold the principles of clarity, transparency, and reproducibility.

Further methodological considerations include the explicit characterization of the analytical approach as either univariate or multivariate. Any preprocessing procedures applied to the selected data, such as filtering, normalization, or other transformations, demand thorough documentation to ensure transparency and facilitate replication. Additionally, the research must identify and rigorously justify the selection of evaluation metrics, ensuring they are appropriately aligned with the specific research task. Ultimately, a comprehensive comparative analysis against established classical and/or state-of-the-art (SOTA) methodologies is indispensable for benchmarking performance and effectively contextualizing the contributions of the proposed work.

\section*{Code availability}


The dataset was generated using open-source software, with all relevant modules publicly available on GitHub. The primary data crawling and processing scripts are hosted at \url{https://github.com/sjtugzx/SpaceTrack-TimeSeries}, enabling full reproducibility and transparency of the data generation workflow.



\bibliography{sample}

\begin{thebibliography}{10}
\urlstyle{rm}
\expandafter\ifx\csname url\endcsname\relax
  \def\url#1{\texttt{#1}}\fi
\expandafter\ifx\csname urlprefix\endcsname\relax\def\urlprefix{URL }\fi
\expandafter\ifx\csname doiprefix\endcsname\relax\def\doiprefix{DOI: }\fi
\providecommand{\bibinfo}[2]{#2}
\providecommand{\eprint}[2][]{\url{#2}}

\bibitem{pachler2021updated}
\bibinfo{author}{Pachler, N.}, \bibinfo{author}{Del~Portillo, I.}, \bibinfo{author}{Crawley, E.~F.} \& \bibinfo{author}{Cameron, B.~G.}
\newblock \bibinfo{title}{An updated comparison of four low earth orbit satellite constellation systems to provide global broadband}.
\newblock In \emph{\bibinfo{booktitle}{2021 IEEE international conference on communications workshops (ICC workshops)}}, \bibinfo{pages}{1--7} (\bibinfo{organization}{IEEE}, \bibinfo{year}{2021}).

\bibitem{shaengchart2024impact}
\bibinfo{author}{Shaengchart, Y.} \& \bibinfo{author}{Kraiwanit, T.}
\newblock \bibinfo{journal}{\bibinfo{title}{An impact of starlink project on regulations and national security}}.
\newblock {\emph{\JournalTitle{Journal of Infrastructure, Policy and Development}}} \textbf{\bibinfo{volume}{8}}, \bibinfo{pages}{3223} (\bibinfo{year}{2024}).

\bibitem{lagunas2024low}
\bibinfo{author}{Lagunas, E.}, \bibinfo{author}{Chatzinotas, S.} \& \bibinfo{author}{Ottersten, B.}
\newblock \bibinfo{journal}{\bibinfo{title}{Low-earth orbit satellite constellations for global communication network connectivity}}.
\newblock {\emph{\JournalTitle{Nature Reviews Electrical Engineering}}} \textbf{\bibinfo{volume}{1}}, \bibinfo{pages}{656--665} (\bibinfo{year}{2024}).

\bibitem{hui2025review}
\bibinfo{author}{Hui, M.} \emph{et~al.}
\newblock \bibinfo{journal}{\bibinfo{title}{A review of leo satellite communication payloads for integrated communication, navigation, and remote sensing: Opportunities, challenges, future directions}}.
\newblock {\emph{\JournalTitle{IEEE Internet of Things Journal}}}  (\bibinfo{year}{2025}).

\bibitem{ai2025research}
\bibinfo{author}{Ai-rong, L.}, \bibinfo{author}{Yong-qing, X.}, \bibinfo{author}{Jian-jiang, H.}, \bibinfo{author}{Xiao-li, X.} \& \bibinfo{author}{Jun, G.}
\newblock \bibinfo{journal}{\bibinfo{title}{Research on starlink ephemeris published by spacex}}.
\newblock {\emph{\JournalTitle{Chinese Astronomy and Astrophysics}}} \textbf{\bibinfo{volume}{49}}, \bibinfo{pages}{193--218} (\bibinfo{year}{2025}).

\bibitem{grile2025statistical}
\bibinfo{author}{Grile, T.~M.}, \bibinfo{author}{Wagenblast, B.~N.} \& \bibinfo{author}{Bettinger, R.~A.}
\newblock \bibinfo{journal}{\bibinfo{title}{Statistical reliability estimation of deep space satellites and launch vehicles: 1958--2022}}.
\newblock {\emph{\JournalTitle{Journal of Spacecraft and Rockets}}} \textbf{\bibinfo{volume}{62}}, \bibinfo{pages}{93--113} (\bibinfo{year}{2025}).

\bibitem{shirobokov2021survey}
\bibinfo{author}{Shirobokov, M.}, \bibinfo{author}{Trofimov, S.} \& \bibinfo{author}{Ovchinnikov, M.}
\newblock \bibinfo{journal}{\bibinfo{title}{Survey of machine learning techniques in spacecraft control design}}.
\newblock {\emph{\JournalTitle{Acta Astronautica}}} \textbf{\bibinfo{volume}{186}}, \bibinfo{pages}{87--97} (\bibinfo{year}{2021}).

\bibitem{caldas2024machine}
\bibinfo{author}{Caldas, F.} \& \bibinfo{author}{Soares, C.}
\newblock \bibinfo{journal}{\bibinfo{title}{Machine learning in orbit estimation: A survey}}.
\newblock {\emph{\JournalTitle{Acta Astronautica}}}  (\bibinfo{year}{2024}).

\bibitem{vallado2001fundamentals}
\bibinfo{author}{Vallado, D.~A.}
\newblock \emph{\bibinfo{title}{Fundamentals of astrodynamics and applications}}, vol.~\bibinfo{volume}{12} (\bibinfo{publisher}{Springer Science \& Business Media}, \bibinfo{year}{2001}).

\bibitem{smith1962application}
\bibinfo{author}{Smith, G.~L.}, \bibinfo{author}{Schmidt, S.~F.} \& \bibinfo{author}{McGee, L.~A.}
\newblock \emph{\bibinfo{title}{Application of statistical filter theory to the optimal estimation of position and velocity on board a circumlunar vehicle}}, vol. \bibinfo{volume}{135} (\bibinfo{publisher}{National Aeronautics and Space Administration}, \bibinfo{year}{1962}).

\bibitem{julier1997new}
\bibinfo{author}{Julier, S.~J.} \& \bibinfo{author}{Uhlmann, J.~K.}
\newblock \bibinfo{title}{New extension of the kalman filter to nonlinear systems}.
\newblock In \emph{\bibinfo{booktitle}{Signal processing, sensor fusion, and target recognition VI}}, vol. \bibinfo{volume}{3068}, \bibinfo{pages}{182--193} (\bibinfo{organization}{Spie}, \bibinfo{year}{1997}).

\bibitem{einicke2012robust}
\bibinfo{author}{Einicke, G.}
\newblock \bibinfo{title}{Robust prediction, filtering and smoothing}.
\newblock In \emph{\bibinfo{booktitle}{Smoothing, Filtering and Prediction-Estimating The Past, Present and Future}} (\bibinfo{publisher}{IntechOpen}, \bibinfo{year}{2012}).

\bibitem{vallado2013improved}
\bibinfo{author}{Vallado, D.~A.}, \bibinfo{author}{Virgili, B.~B.} \& \bibinfo{author}{Flohrer, T.}
\newblock \bibinfo{title}{Improved ssa through orbit determination of two-line element sets}.
\newblock In \emph{\bibinfo{booktitle}{ESA Space Debris Conference}} (\bibinfo{year}{2013}).

\bibitem{miura2009comparison}
\bibinfo{author}{Miura, N.~Z.}
\newblock \emph{\bibinfo{title}{Comparison and design of Simplified General Perturbation Models (SGP4) and code for NASA Johnson Space Center, Orbital debris program office}} (\bibinfo{publisher}{California Polytechnic State University}, \bibinfo{year}{2009}).

\bibitem{urrutxua2016dromo}
\bibinfo{author}{Urrutxua, H.}, \bibinfo{author}{Sanjurjo-Rivo, M.} \& \bibinfo{author}{Pel{\'a}ez, J.}
\newblock \bibinfo{journal}{\bibinfo{title}{Dromo propagator revisited}}.
\newblock {\emph{\JournalTitle{Celestial Mechanics and Dynamical Astronomy}}} \textbf{\bibinfo{volume}{124}}, \bibinfo{pages}{1--31} (\bibinfo{year}{2016}).

\bibitem{sharma1988long}
\bibinfo{author}{Sharma, R.~K.} \& \bibinfo{author}{James~Raj, M.}
\newblock \bibinfo{journal}{\bibinfo{title}{Long-term orbit computations with ks uniformly regular canonical elements with oblateness}}.
\newblock {\emph{\JournalTitle{Earth, Moon, and Planets}}} \textbf{\bibinfo{volume}{42}}, \bibinfo{pages}{163--178} (\bibinfo{year}{1988}).

\bibitem{aristoff2014orbit}
\bibinfo{author}{Aristoff, J.~M.}, \bibinfo{author}{Horwood, J.~T.} \& \bibinfo{author}{Poore, A.~B.}
\newblock \bibinfo{journal}{\bibinfo{title}{Orbit and uncertainty propagation: a comparison of gauss--legendre-, dormand--prince-, and chebyshev--picard-based approaches}}.
\newblock {\emph{\JournalTitle{Celestial Mechanics and Dynamical Astronomy}}} \textbf{\bibinfo{volume}{118}}, \bibinfo{pages}{13--28} (\bibinfo{year}{2014}).

\bibitem{bai2011modified}
\bibinfo{author}{Bai, X.} \& \bibinfo{author}{Junkins, J.~L.}
\newblock \bibinfo{journal}{\bibinfo{title}{Modified chebyshev-picard iteration methods for orbit propagation}}.
\newblock {\emph{\JournalTitle{The Journal of the Astronautical Sciences}}} \textbf{\bibinfo{volume}{58}}, \bibinfo{pages}{583--613} (\bibinfo{year}{2011}).

\bibitem{bradley2012new}
\bibinfo{author}{Bradley, B.~K.}, \bibinfo{author}{Jones, B.~A.}, \bibinfo{author}{Beylkin, G.} \& \bibinfo{author}{Axelrad, P.}
\newblock \bibinfo{title}{A new numerical integration technique in astrodynamics}.
\newblock In \emph{\bibinfo{booktitle}{22nd AAS/AIAA Space Flight Mechanics Meeting, Charleston, SC, AAS}}, \bibinfo{pages}{12--216} (\bibinfo{year}{2012}).

\bibitem{levit2011improved}
\bibinfo{author}{Levit, C.} \& \bibinfo{author}{Marshall, W.}
\newblock \bibinfo{journal}{\bibinfo{title}{Improved orbit predictions using two-line elements}}.
\newblock {\emph{\JournalTitle{Advances in Space Research}}} \textbf{\bibinfo{volume}{47}}, \bibinfo{pages}{1107--1115} (\bibinfo{year}{2011}).

\bibitem{bennett2012improving}
\bibinfo{author}{Bennett, J.}, \bibinfo{author}{Sang, J.}, \bibinfo{author}{Smith, C.} \& \bibinfo{author}{Zhang, K.}
\newblock \bibinfo{title}{Improving low-earth orbit predictions using two-line element data with bias correction}.
\newblock In \emph{\bibinfo{booktitle}{Advanced Maui Optical and Space Surveillance Technologies Conference}}, vol.~\bibinfo{volume}{1}, \bibinfo{pages}{46} (\bibinfo{year}{2012}).

\bibitem{sang2017analytical}
\bibinfo{author}{Sang, J.}, \bibinfo{author}{Li, B.}, \bibinfo{author}{Chen, J.}, \bibinfo{author}{Zhang, P.} \& \bibinfo{author}{Ning, J.}
\newblock \bibinfo{journal}{\bibinfo{title}{Analytical representations of precise orbit predictions for earth orbiting space objects}}.
\newblock {\emph{\JournalTitle{Advances in Space Research}}} \textbf{\bibinfo{volume}{59}}, \bibinfo{pages}{698--714} (\bibinfo{year}{2017}).

\bibitem{san2017hybrid}
\bibinfo{author}{San-Juan, J.~F.}, \bibinfo{author}{P{\'e}rez, I.}, \bibinfo{author}{San-Mart{\'\i}n, M.} \& \bibinfo{author}{Vergara, E.~P.}
\newblock \bibinfo{journal}{\bibinfo{title}{Hybrid sgp4 orbit propagator}}.
\newblock {\emph{\JournalTitle{Acta Astronautica}}} \textbf{\bibinfo{volume}{137}}, \bibinfo{pages}{254--260} (\bibinfo{year}{2017}).

\bibitem{peng2020machine}
\bibinfo{author}{Peng, H.} \& \bibinfo{author}{Bai, X.}
\newblock \bibinfo{journal}{\bibinfo{title}{Machine learning approach to improve satellite orbit prediction accuracy using publicly available data}}.
\newblock {\emph{\JournalTitle{The Journal of the astronautical sciences}}} \textbf{\bibinfo{volume}{67}}, \bibinfo{pages}{762--793} (\bibinfo{year}{2020}).

\bibitem{muldoon2009improved}
\bibinfo{author}{Muldoon, A.~R.}, \bibinfo{author}{Elkaim, G.~H.}, \bibinfo{author}{Rickard, I.~F.} \& \bibinfo{author}{Weeden, B.}
\newblock \bibinfo{journal}{\bibinfo{title}{Improved orbital debris trajectory estimation based on sequential tle processing}}.
\newblock {\emph{\JournalTitle{Paper IAC-09 A}}} \textbf{\bibinfo{volume}{6}} (\bibinfo{year}{2009}).

\bibitem{peng2018exploring}
\bibinfo{author}{Peng, H.} \& \bibinfo{author}{Bai, X.}
\newblock \bibinfo{journal}{\bibinfo{title}{Exploring capability of support vector machine for improving satellite orbit prediction accuracy}}.
\newblock {\emph{\JournalTitle{Journal of aerospace information systems}}} \textbf{\bibinfo{volume}{15}}, \bibinfo{pages}{366--381} (\bibinfo{year}{2018}).

\bibitem{peng2019comparative}
\bibinfo{author}{Peng, H.} \& \bibinfo{author}{Bai, X.}
\newblock \bibinfo{journal}{\bibinfo{title}{Comparative evaluation of three machine learning algorithms on improving orbit prediction accuracy}}.
\newblock {\emph{\JournalTitle{Astrodynamics}}} \textbf{\bibinfo{volume}{3}}, \bibinfo{pages}{325--343} (\bibinfo{year}{2019}).

\bibitem{peng2019gaussian}
\bibinfo{author}{Peng, H.} \& \bibinfo{author}{Bai, X.}
\newblock \bibinfo{journal}{\bibinfo{title}{Gaussian processes for improving orbit prediction accuracy}}.
\newblock {\emph{\JournalTitle{Acta astronautica}}} \textbf{\bibinfo{volume}{161}}, \bibinfo{pages}{44--56} (\bibinfo{year}{2019}).

\bibitem{peng2021fusion}
\bibinfo{author}{Peng, H.} \& \bibinfo{author}{Bai, X.}
\newblock \bibinfo{journal}{\bibinfo{title}{Fusion of a machine learning approach and classical orbit predictions}}.
\newblock {\emph{\JournalTitle{Acta astronautica}}} \textbf{\bibinfo{volume}{184}}, \bibinfo{pages}{222--240} (\bibinfo{year}{2021}).

\bibitem{rautalin2017latent}
\bibinfo{author}{Rautalin, S.}, \bibinfo{author}{Ali-L{\"o}ytty, S.} \& \bibinfo{author}{Pich{\'e}, R.}
\newblock \bibinfo{title}{Latent force models in autonomous gnss satellite orbit prediction}.
\newblock In \emph{\bibinfo{booktitle}{2017 International Conference on Localization and GNSS (ICL-GNSS)}}, \bibinfo{pages}{1--6} (\bibinfo{organization}{IEEE}, \bibinfo{year}{2017}).

\bibitem{li2021improved}
\bibinfo{author}{Li, B.}, \bibinfo{author}{Zhang, Y.}, \bibinfo{author}{Huang, J.} \& \bibinfo{author}{Sang, J.}
\newblock \bibinfo{journal}{\bibinfo{title}{Improved orbit predictions using two-line elements through error pattern mining and transferring}}.
\newblock {\emph{\JournalTitle{Acta Astronautica}}} \textbf{\bibinfo{volume}{188}}, \bibinfo{pages}{405--415} (\bibinfo{year}{2021}).

\bibitem{pihlajasalo2018improvement}
\bibinfo{author}{Pihlajasalo, J.}, \bibinfo{author}{Lepp{\"a}koski, H.}, \bibinfo{author}{Ali-L{\"o}ytty, S.} \& \bibinfo{author}{Piche, R.}
\newblock \bibinfo{title}{Improvement of gps and beidou extended orbit predictions with cnns}.
\newblock In \emph{\bibinfo{booktitle}{2018 European Navigation Conference (ENC)}}, \bibinfo{pages}{54--59} (\bibinfo{organization}{IEEE}, \bibinfo{year}{2018}).

\bibitem{san2018hybrid}
\bibinfo{author}{San-Juan, J.~F.} \emph{et~al.}
\newblock \bibinfo{title}{Hybrid sgp4 propagator based on machine-learning techniques applied to galileo-type orbits}.
\newblock In \emph{\bibinfo{booktitle}{69th International Astronautical Congress, Bremen, Germany}}, vol.~\bibinfo{volume}{4} (\bibinfo{year}{2018}).

\bibitem{curzi2022two}
\bibinfo{author}{Curzi, G.}, \bibinfo{author}{Modenini, D.} \& \bibinfo{author}{Tortora, P.}
\newblock \bibinfo{journal}{\bibinfo{title}{Two-line-element propagation improvement and uncertainty estimation using recurrent neural networks}}.
\newblock {\emph{\JournalTitle{CEAS Space Journal}}} \textbf{\bibinfo{volume}{14}}, \bibinfo{pages}{197--204} (\bibinfo{year}{2022}).

\bibitem{salleh2021adaptation}
\bibinfo{author}{Salleh, N.}, \bibinfo{author}{Azmi, N. F.~M.} \& \bibinfo{author}{Yuhaniz, S.~S.}
\newblock \bibinfo{title}{An adaptation of deep learning technique in orbit propagation model using long short-term memory}.
\newblock In \emph{\bibinfo{booktitle}{2021 International Conference on Electrical, Communication, and Computer Engineering (ICECCE)}}, \bibinfo{pages}{1--6} (\bibinfo{organization}{IEEE}, \bibinfo{year}{2021}).

\bibitem{li2020machine}
\bibinfo{author}{Li, B.}, \bibinfo{author}{Huang, J.}, \bibinfo{author}{Feng, Y.}, \bibinfo{author}{Wang, F.} \& \bibinfo{author}{Sang, J.}
\newblock \bibinfo{journal}{\bibinfo{title}{A machine learning-based approach for improved orbit predictions of leo space debris with sparse tracking data from a single station}}.
\newblock {\emph{\JournalTitle{IEEE Transactions on Aerospace and Electronic Systems}}} \textbf{\bibinfo{volume}{56}}, \bibinfo{pages}{4253--4268} (\bibinfo{year}{2020}).

\bibitem{tipaldi2022reinforcement}
\bibinfo{author}{Tipaldi, M.}, \bibinfo{author}{Iervolino, R.} \& \bibinfo{author}{Massenio, P.~R.}
\newblock \bibinfo{journal}{\bibinfo{title}{Reinforcement learning in spacecraft control applications: Advances, prospects, and challenges}}.
\newblock {\emph{\JournalTitle{Annual Reviews in Control}}} \textbf{\bibinfo{volume}{54}}, \bibinfo{pages}{1--23} (\bibinfo{year}{2022}).

\bibitem{blasch2022space}
\bibinfo{author}{Blasch, E.}, \bibinfo{author}{Chen, G.}, \bibinfo{author}{Shen, D.}, \bibinfo{author}{Insaurralde, C.~C.} \& \bibinfo{author}{Pham, K.}
\newblock \bibinfo{title}{Space track ontology elements for space domain awareness}.
\newblock In \emph{\bibinfo{booktitle}{Sensors and Systems for Space Applications XV}}, vol. \bibinfo{volume}{12121}, \bibinfo{pages}{98--107} (\bibinfo{organization}{SPIE}, \bibinfo{year}{2022}).

\bibitem{liu2024maneuver}
\bibinfo{author}{Liu, A.}, \bibinfo{author}{Xu, X.}, \bibinfo{author}{Xiong, Y.} \& \bibinfo{author}{Yu, S.}
\newblock \bibinfo{journal}{\bibinfo{title}{Maneuver strategies of starlink satellite based on spacex-released ephemeris}}.
\newblock {\emph{\JournalTitle{Advances in Space Research}}} \textbf{\bibinfo{volume}{74}}, \bibinfo{pages}{3157--3169} (\bibinfo{year}{2024}).

\bibitem{capderou2005satellite}
\bibinfo{author}{Capderou, M.}
\newblock \bibinfo{journal}{\bibinfo{title}{Satellite in keplerian orbit}}.
\newblock {\emph{\JournalTitle{Satellites: Orbits and Missions}}} \bibinfo{pages}{41--58} (\bibinfo{year}{2005}).

\bibitem{kelso2017challenges}
\bibinfo{author}{Kelso, T.}
\newblock \bibinfo{title}{Challenges identifying newly launched objects}.
\newblock In \emph{\bibinfo{booktitle}{68th International Astronautical Congress, Adelaide, Australia}} (\bibinfo{year}{2017}).

\bibitem{harris2019accurate}
\bibinfo{author}{Harris, B.}
\newblock \bibinfo{title}{Accurate local timestamps}.
\newblock In \emph{\bibinfo{booktitle}{Proceedings of the 50th Annual Precise Time and Time Interval Systems and Applications Meeting}}, \bibinfo{pages}{364--373} (\bibinfo{year}{2019}).

\bibitem{guo2025spacetrack}
\bibinfo{author}{Guo, Z.} \emph{et~al.}
\newblock \bibinfo{title}{{SpaceTrack-TimeSeries: Time Series Dataset towards Satellite Orbit Analysis}}.
\newblock \bibinfo{howpublished}{Dataset}, \url{10.6084/m9.figshare.29067764.v1} (\bibinfo{year}{2025}).

\bibitem{dong2010accuracy}
\bibinfo{author}{Dong, W.} \& \bibinfo{author}{Chang-yin, Z.}
\newblock \bibinfo{journal}{\bibinfo{title}{An accuracy analysis of the sgp4/sdp4 model}}.
\newblock {\emph{\JournalTitle{Chinese Astronomy and Astrophysics}}} \textbf{\bibinfo{volume}{34}}, \bibinfo{pages}{69--76} (\bibinfo{year}{2010}).

\bibitem{lanz2007grid}
\bibinfo{author}{Lanz, T.} \& \bibinfo{author}{Hubeny, I.}
\newblock \bibinfo{journal}{\bibinfo{title}{A grid of nlte line-blanketed model atmospheres of early b-type stars}}.
\newblock {\emph{\JournalTitle{The Astrophysical Journal Supplement Series}}} \textbf{\bibinfo{volume}{169}}, \bibinfo{pages}{83} (\bibinfo{year}{2007}).

\bibitem{peloni2018osculating}
\bibinfo{author}{Peloni, A.}, \bibinfo{author}{McInnes, C.~R.} \& \bibinfo{author}{Ceriotti, M.}
\newblock \bibinfo{journal}{\bibinfo{title}{Osculating keplerian elements for highly non-keplerian orbits}}.
\newblock {\emph{\JournalTitle{Journal of Guidance, Control, and Dynamics}}} \textbf{\bibinfo{volume}{41}}, \bibinfo{pages}{2489--2498} (\bibinfo{year}{2018}).

\bibitem{marsouin1991navigation}
\bibinfo{author}{Marsouin, A.} \& \bibinfo{author}{Brunel, P.}
\newblock \bibinfo{journal}{\bibinfo{title}{Navigation of avhrr images using argos or tbus bulletins}}.
\newblock {\emph{\JournalTitle{International Journal of Remote Sensing}}} \textbf{\bibinfo{volume}{12}}, \bibinfo{pages}{1575--1592} (\bibinfo{year}{1991}).

\bibitem{ude2014orientation}
\bibinfo{author}{Ude, A.}, \bibinfo{author}{Nemec, B.}, \bibinfo{author}{Petri{\'c}, T.} \& \bibinfo{author}{Morimoto, J.}
\newblock \bibinfo{title}{Orientation in cartesian space dynamic movement primitives}.
\newblock In \emph{\bibinfo{booktitle}{2014 IEEE International Conference on Robotics and Automation (ICRA)}}, \bibinfo{pages}{2997--3004} (\bibinfo{organization}{IEEE}, \bibinfo{year}{2014}).

\end{thebibliography}

\section*{Author contributions statement}
Zhixin Guo led conceptualization, methodology design, software development, data validation, data analysis, writing ( original draft), and contributed to writing (review \& editing). Qi Shi contributed to conceptualization and writing (review \& editing). Xiaofan Xu led project administration and supervision, and contributed to conceptualization and writing (review \& editing). Sixiang Shan led data crawling, and contributed to writing (review \& editing). Limin Qin contributed to data validation and writing (review \& editing). Linqiang Ge and Rui Zhang contributed to project administration and supervision. Ya Dai, Hua Zhu, and Guowei Jiang each contributed to conceptualization and writing (review \& editing).

\section*{Competing interests}
The authors declare no competing interests.

\end{document}